\documentclass{epl}
\usepackage{epsfig}
\usepackage{amssymb,amsbsy,amsmath}
\newcommand{\op}[1]{%
    \fontdimen12\textfont3=2pt\fontdimen12\scriptfont3=1.4pt%
    \!\null\mathop{\vphantom{#1}\smash{#1}}\limits_{\sim}\null\!}
\newcommand{\vek}[1]{{\!\vec{\,#1}}}

\newcommand{\figref}[1]{Fig.~\protect\ref{#1}}
\newcommand{\fmref}[1]{(\protect\ref{#1})}

\def\geap{\raisebox{-.6ex}{$\stackrel {>}{\sim}$}} 
\title{Quantum rotational band model for the
Heisenberg molecular magnet
\{Mo$_{72}$Fe$_{30}$\}} 
\shorttitle{Rotational band model}
\author{J\"urgen Schnack\inst{1}, Marshall Luban\inst{2}, \and
Robert Modler\inst{2}}
\shortauthor{J.~Schnack \and M.~Luban \and R.~Modler}
\institute{
  \inst{1} Universit\"at Osnabr\"uck, Fachbereich Physik,
D-49069 Osnabr\"uck, Germany\\
  \inst{2} Ames Laboratory \& Department of Physics and Astronomy,
Iowa State University, Ames, Iowa 50011, USA
}

\begin{document}

\maketitle

\begin{abstract}
We derive the low temperature properties of the molecular magnet
\{Mo$_{72}$Fe$_{30}$\}, where 30 Fe$^{3+}$ paramagnetic ions
occupy the sites of an icosidodecahedron and interact via
isotropic nearest-neighbour antiferromagnetic Heisenberg
exchange. The key idea of our model (J.S. \& M.L.) is that the
low-lying excitations form a sequence of ``rotational bands",
i.e., for each such band the excitation energies depend
quadratically on the total spin quantum number. For temperatures
below 50~mK we predict that the magnetisation is described by a
staircase with 75 equidistant steps as the magnetic field is
increased up to a critical value and saturated for higher
fields. For higher temperatures thermal broadening effects wash
out the staircase and yield a linear ramp below the critical
field, and this has been confirmed by our measurements
(R.M.). We demonstrate that the lowest two rotational bands are
separated by an energy gap of 0.7~meV, and this could be tested
by EPR and inelastic neutron scattering measurements. We also
predict the occurrence of resonances at temperatures below 0.1~K
in the proton NMR spin-lattice relaxation rate associated with
level crossings.  As rotational bands characterize the spectra
of many magnetic molecules our method opens a new road towards a
description of their low-temperature behaviour which is not
otherwise accessible.
\end{abstract}

\section{Introduction}
A new class of magnetic compounds known as molecular magnets
\cite{r-1} is attracting much attention. These compounds can be
synthesized as single crystals of identical molecular units,
each containing several paramagnetic ions that mutually interact
via Heisenberg exchange. The intermolecular magnetic
interactions are in the great majority of cases utterly
negligible as compared to intramolecular magnetic
interactions. Measurements of the magnetic properties therefore
reflect those of the common, individual molecular
unit. Molecular magnets such as \{Mn$_{12}$\} and \{Fe$_{8}$\}
have been the focal point for intense study of subjects of broad
scientific importance, such as quantum tunneling of
magnetisation and quantum coherence \cite{r-2}. However, it is
doubtful \cite{r-3} that larger symmetric arrays of paramagnetic
ions can be accommodated in such polynuclear coordination
complexes utilizing simple organic bridging ligands. Very
recently the first examples of a new paradigm of molecular
magnets, based on so-called Keplerate structures, have been
synthesized \cite{MSS:ACIE99}, and these offer numerous avenues
for obtaining truly giant, highly symmetric arrays of
paramagnetic ions. The archetype of this new class is referred
to as \{Mo$_{72}$Fe$_{30}$\} \cite{MSS:ACIE99}. Embedded within
a (diamagnetic) host molecule (mol. wt.  18,649), 30 Fe$^{3+}$
paramagnetic ions (spins $s=5/2$) occupy the sites of an
icosidodecahedron and interact via isotropic, nearest-neighbour
antiferromagnetic exchange. This Keplerate and its interlinked
derivatives \cite{MSS:SSS00} pose a major theoretical
challenge. The dimension of the Hilbert space for
\{Mo$_{72}$Fe$_{30}$\} is a staggering $6^{30}$, precluding the
calculation of the energy eigenvalues on any computer.

In this Letter we show that despite this immense obstacle the
major low-temperature properties of \{Mo$_{72}$Fe$_{30}$\} can
be calculated.  Rather than pursue the futile task of
diagonalizing the Heisenberg Hamiltonian [see
\fmref{E-1}], we adopt an approximate, diagonalizable Hamilton
operator [see \fmref{E-2}] which properly incorporates the
generic result that the low-lying magnetic energy levels of a
wide class of molecular magnets are arranged as parallel
rotational bands. These rotational bands reflect the underlying
sublattice structure of the spin array \cite{BLL:PRB94,ScL:PRB}
that has been found for the corresponding classical system
\cite{AxL}.  But in contrast to the classical model and
approximations like high-temperature series expansions the
present approach enables us to access the properties of giant
magnetic molecules for very low temperatures.  We are also able
to show that the predictions of the classical model extend to
temperatures as low as 100~mK with very good accuracy for the
special case of \{Mo$_{72}$Fe$_{30}$\}.  In particular, we
present our theoretical predictions for the temperature and
field dependence of the magnetisation, and for the resonances of
the spin-lattice relaxation rate of \{Mo$_{72}$Fe$_{30}$\}. We
also show that our experimental findings directly reflect the
existence of rotational bands.

\section{Rotational band Hamiltonian}

The Hamiltonian for the isotropic Heisenberg model
including the interaction with an external magnetic field $B$
reads
\begin{eqnarray}
\label{E-1}
\op{H}
&=&
-
2\,J\,
\sum_{(u,v)}\;
\op{\vek{s}}(u) \cdot \op{\vek{s}}(v)
+
g \mu_B B \op{S}_z
\ ,
\end{eqnarray}
where $J$ is the exchange constant with units of energy, and
$J<0$ corresponds to antiferromagnetic coupling, $g$ is the
spectroscopic splitting factor, and $\mu_B$ is the Bohr
magneton. The vector operators $\op{\vek{s}}(u)$ are the spin
operators (in units of $\hbar$) of the individual paramagnetic
ions with spin $s$. The sum in \eqref{E-1} runs over all
distinct nearest-neighbour pairs $(u,v)$ of spins of a single
molecule at positions $u$ and $v$. Since the Hamiltonian
commutes with the operators $\op{\vek{S}}^2$ and $\op{S}_z$ of
the total spin, the eigenstates of $\op{H}$ may be classified
using the quantum numbers $S$ and $M$.

In order to develop an approximate quantum model of
\{Mo$_{72}$Fe$_{30}$\}, we first exploit the fact that the set
of minimal energies for each $S$ forms a rotational band,
\begin{eqnarray}
\label{E-4}
E_{S, min}
\approx
- J\, [D(N,s)/N]\,
S (S+1)
+
E_a
\ .
\end{eqnarray}
This has been noted on several occasions for ring structures
with an even number $N$ of sites \cite{CCF:CEJ96,ACC:ICA00}.
Elsewhere we reported that all finite Heisenberg systems with
isotropic and homogeneous antiferromagnetic exchange (including
rings for both even and odd $N$, tetrahedron, cube, octahedron,
icosahedron, triangular prism, and axially truncated
icosahedron) exhibit a rotational band \cite{ScL:PRB}.  For
high-symmetry systems we provided an expression for the
classical limit $D(N,\infty)$, henceforth denoted by $D$, which
yields $D=4$ for rings with even $N$, see also \cite{ACC:ICA00},
and $D=6$ for the icosidodecahedron, cube, and octahedron. Our
investigations \cite{ScL:PRB} have shown that the numerical
value of $D(N,s)$ for any finite $s$ is always a little larger
than $D$. This difference can be taken as a measure of quantum
spin effects; consistently it is largest for $s=1/2$.

The constant offset $E_a$ in \fmref{E-4} is to be selected so
that the highest level of the rotational band, which occurs for
$S=N s$, agrees with the largest energy eigenvalue of
\fmref{E-1}. The corresponding eigenvector is the ground state
eigenvector for the counterpart ferromagnetic system. Thus the
largest energy eigenvalue serves as an anchor in all
approximations of the rotational band \cite{ScL:PRB}.
Summarizing, for the molecular magnet \{Mo$_{72}$Fe$_{30}$\} the
lowest rotational band is given by
\begin{eqnarray}
\label{E-5}
E_{S, min}
=
-\frac{J}{5}\,
S (S+1) 
+
60 J s \left(s+\frac{1}{10}\right)
\ .
\end{eqnarray}
\begin{figure}[ht!]
\begin{center}
\epsfig{file=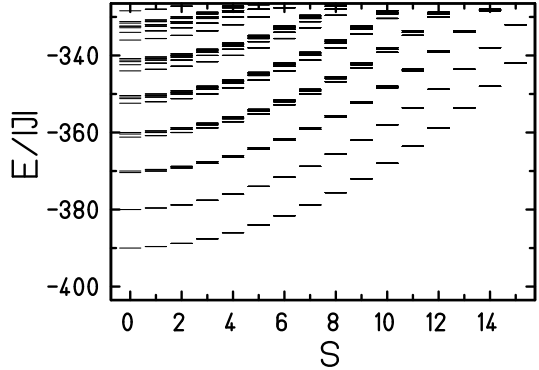,width=70mm}
\vspace*{1mm}
\caption[]{Low-lying energy eigenvalues
using the approximate Hamiltonian \fmref{E-2} with $B=0$. The
levels of the lowest band are given by \eqref{E-5}.}
\label{F-1}
\end{center} 
\end{figure} 

The rotational band \fmref{E-5} by itself is insufficient for
calculating observables at low temperatures; it is essential to
include the other low-lying excitation energies.  This can be
achieved via the introduction of an effective Hamiltonian
whose form is set by incorporating the known symmetry of the
spin array (icosidodecahedron), which is similar to that of the
triangular lattice. Therefore, the low-lying spectrum can be
understood as originating from three interacting sublattice
spins \cite{BLL:PRB94}. These findings are supported by the fact
that also the exact classical ground state of the
icosidodecahedron for $J<0$ is describable in terms of three
sublattice spin vectors ${\vek{S}}_{A}$, ${\vek{S}}_{B}$, and
${\vek{S}}_{C}$ with $S_{A}=S_{B}=S_{C}=25$, and with relative
angles of $120^o$ \cite{AxL}. We thus adopt as our effective
Hamiltonian, replacing the field-free term of \fmref{E-1},
see also \cite{ScL:PRB,BLL:PRB94,HeZ:PRL81},
\begin{eqnarray}
\label{E-2}
\op{H}_1^{\text{eff}}
&=&
-\frac{J}{5}\,
\big[
\op{\vek{S}}^2 
- 
\left(
\op{\vek{S}}_A^2 + \op{\vek{S}}_B^2 + \op{\vek{S}}_C^2
\right)
\big]
\ .
\end{eqnarray}
The three sublattice spin quantum numbers, $S_A, S_B, S_C$, can
assume values $0, 1,\dots, 25$. The sublattice spin operators
mutually commute and they also commute with
$\op{H}_1^{\text{eff}}$.  Thus the eigenvalues of
$\op{H}_1^{\text{eff}}$ are given in terms of the quantum
numbers $S$, $S_{A}$, $S_{B}$, and $S_{C}$. The lowest
rotational band, \eqref{E-5}, arises upon choosing
$S_{A}=S_{B}=S_{C}=25$ and allowing $S$ to extend from 0 to
75. The next higher rotational band, see \figref{F-1}, is
obtained for the choice $S_{A}=S_{B}=25$, $S_{C}=24$ and its
permutations. Note that these two bands are separated by an
energy gap $\Delta=10 J$. Continuing this process leads to a
sequence of parabolic bands. This is rather realistic for the
second band, and indeed observed in many finite Heisenberg
systems, compare Figs. 1, 3, and 4 in \cite{ScL:PRB}.

\section{Experimental implications}

The theoretical result for ${\mathcal M}/(g \mu_B)$ versus $B$
at $T=0$~K, as obtained using \fmref{E-2} and a Zeeman term, is
a staircase with $75$ steps of unit height which terminates at
the critical field $B_c=30|J|/(g \mu_B)=17.7$~T. The values of
$J$ and $g$ have been determined by high-temperature, low-field
magnetic susceptibility measurements \cite{CPC}:
$J/k_B=-0.783$~K, and $g=1.974$. For $B>B_c$ all spins are
aligned parallel and the total moment is given by ${\mathcal
M}=75g \mu_B$. 

\begin{figure}[ht!]
\begin{center}
\epsfig{file=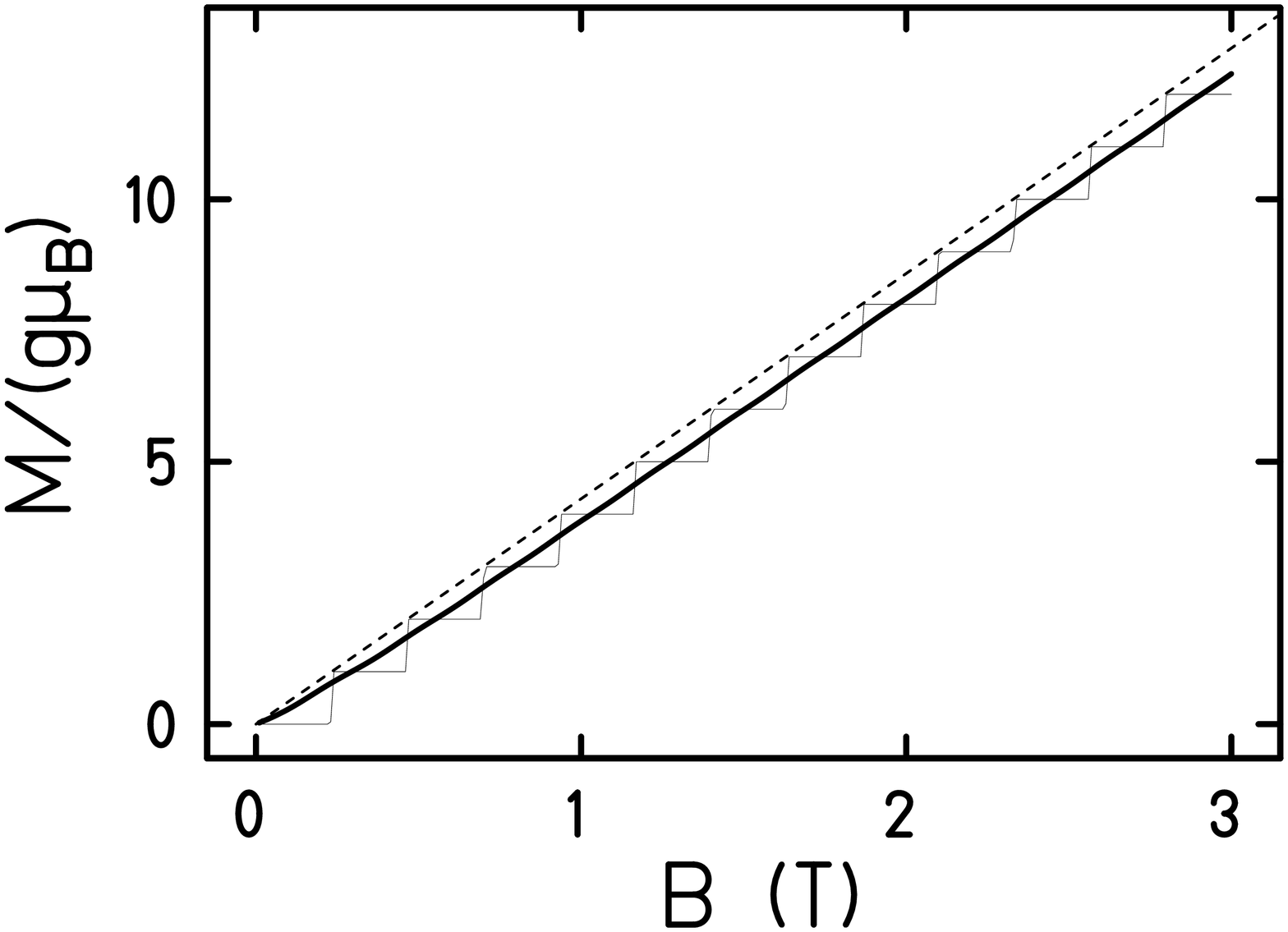,width=65mm}
$\quad$
\epsfig{file=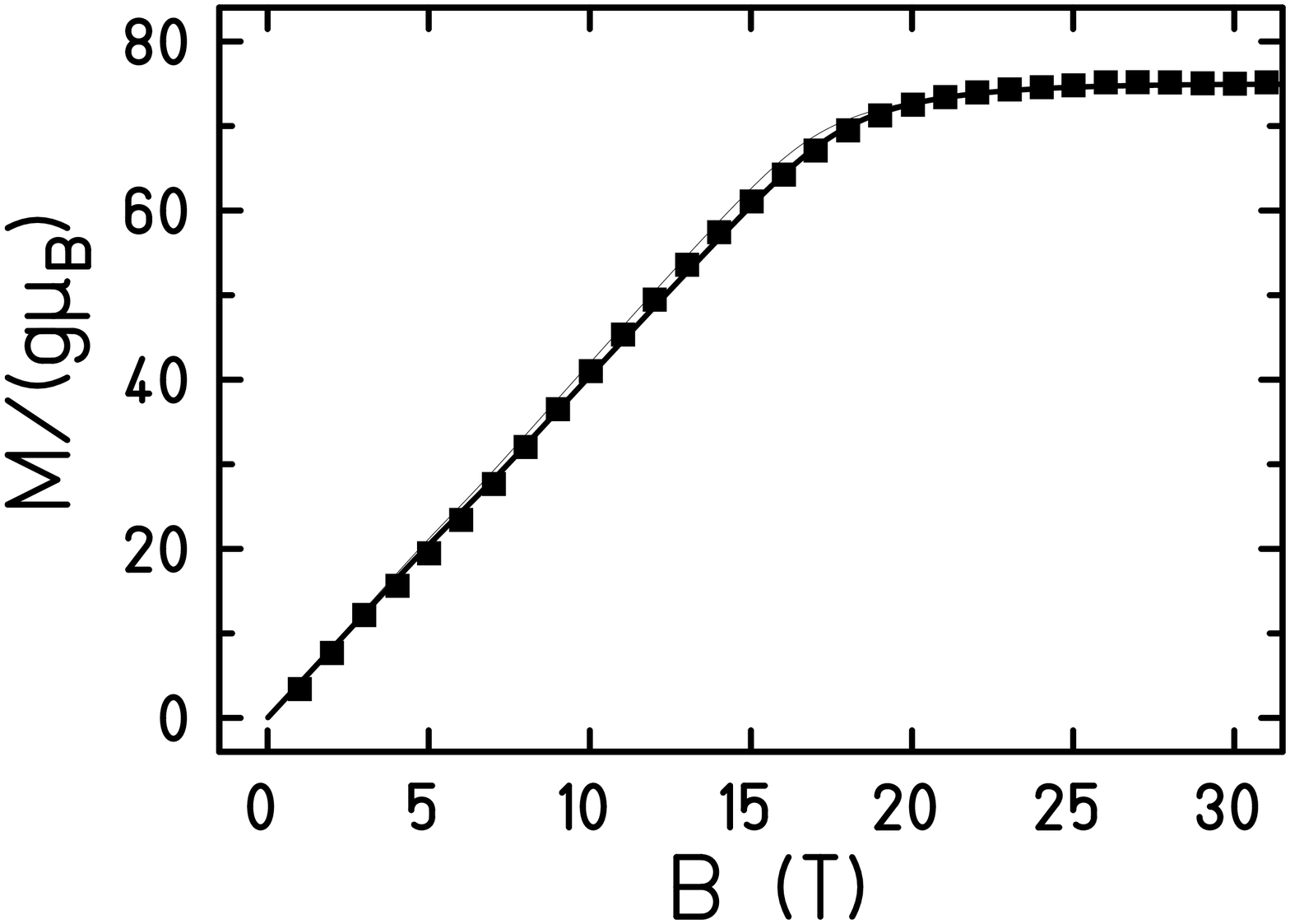,width=65mm}
\vspace*{1mm}
\caption[]{L.h.s.: The magnetisation according to \fmref{E-2} is
shown for $T=1$~mK (staircase) and 100~mK (thick solid curve) as
well as for the corresponding classical Heisenberg model at
$T=0$ (dashed curve).  R.h.s.: The thin curve displays the
magnetisation following from \fmref{E-2} for
$T=4$~K. Experimental data using a pulsed field are given by the
solid squares, their size reflects an uncertainty of $\pm 0.5$~T
for the data. The thick curve is gives the result of the
improved approximation \fmref{E-3}, also at 4~K, and closely
reproduces the measured values.}
\label{F-2}
\end{center} 
\end{figure} 
In the left panel of \figref{F-2} we provide a graph of
${\mathcal M}/(g \mu_B)$ versus $B$ as predicted by the quantum
model for $T=1$~mK (staircase) and 100~mK (thick solid curve),
as well as the rigorous result for the corresponding classical
Heisenberg model at $T=0$, a strictly linear dependence of M on
$B$ up to $B_c$ (dashed curve).  In fact, for $T>50$~mK thermal
broadening effects smear out the staircase of the quantum model
and, above approximately $B=0.5$~T, ${\mathcal M}$ increases
linearly with $B$, almost parallel to but somewhat below the
classical $T=0$ result. It would be extremely interesting to put
this prediction of staircase-like behavior below 50~mK to a
careful experimental test. 

What is currently available are experimental data for ${\mathcal
M}$ versus $B$ on \{Mo$_{72}$Fe$_{30}$\}, see the right panel of
\figref{F-2}, as obtained using a pulsed magnetic field. Due to
the rapid introduction of the magnetic field, increasing from 0
to 60 Tesla in approximately 6~ms, we estimate that the
effective temperature of the spins was 4~K, even though the
nominal cryostat temperature was 0.46~K.  Nevertheless, the very
good agreement between theory and experiment confirms the
underlying picture of rotational bands on which the model is
based. For a non-parabolic dependence of the energy eigenvalues
on $S$ the magnetisation curve would show unequal steps at $T=0$
as well as nonlinear behaviour at higher temperatures.  As
remarked above, the classical coefficient $D=D(N,\infty)$ always
underestimates the true coefficient $D(N,s)$ by a few percent
\cite{ScL:PRB}.  Of course, there is no simple method for
establishing the correct value of $D(30,5/2)$, so it is of
interest to estimate its value by using the currently available
experimental data for ${\mathcal M}$ versus $B$.  We attempt to
improve approximation \fmref{E-2} quantitatively while ensuring
that the rotational band structure is not altered. We adjust
$D(N,s)$ so that the resulting magnetisation curve (thick curve
in \figref{F-2}, r.h.s.) provides an optimal fit to the measured
data. This is achieved by taking $D(N,s)=6.23$, which is very
close to $D=6$. Thus an alternate effective Hamiltonian is given
by
\begin{eqnarray}
\label{E-3}
\op{H}_2^{\text{eff}}
&=&
-J\, \frac{D(N,s)}{N}\,
\left[
\op{\vek{S}}^2 
- 
\gamma
\left(
\op{\vek{S}}_A^2 + \op{\vek{S}}_B^2 + \op{\vek{S}}_C^2
\right)
\right]
\ ,
\end{eqnarray}
with $\gamma=1.07$ in order to maintain the correct value of
the largest energy eigenvalue. Using \fmref{E-3} we find that
$B_c\approx 18.4$~T. Finally, using this tentative estimate for
$D(30,5/2)$ enables us to predict the ground state energy of
\{Mo$_{72}$Fe$_{30}$\} as $E_0/k_B\approx -339$~K. Taking into
account the error bars of the measured data we estimate that the
resulting uncertainty of $D(30,5/2)$, $\gamma$, and $E_0$ is less
than 3~\%. We are currently attempting to calculate $D(30,5/2)$
from first principles using DMRG techniques \cite{Whi:PRB93}.

\begin{figure}[ht!]
\begin{center}
\epsfig{file=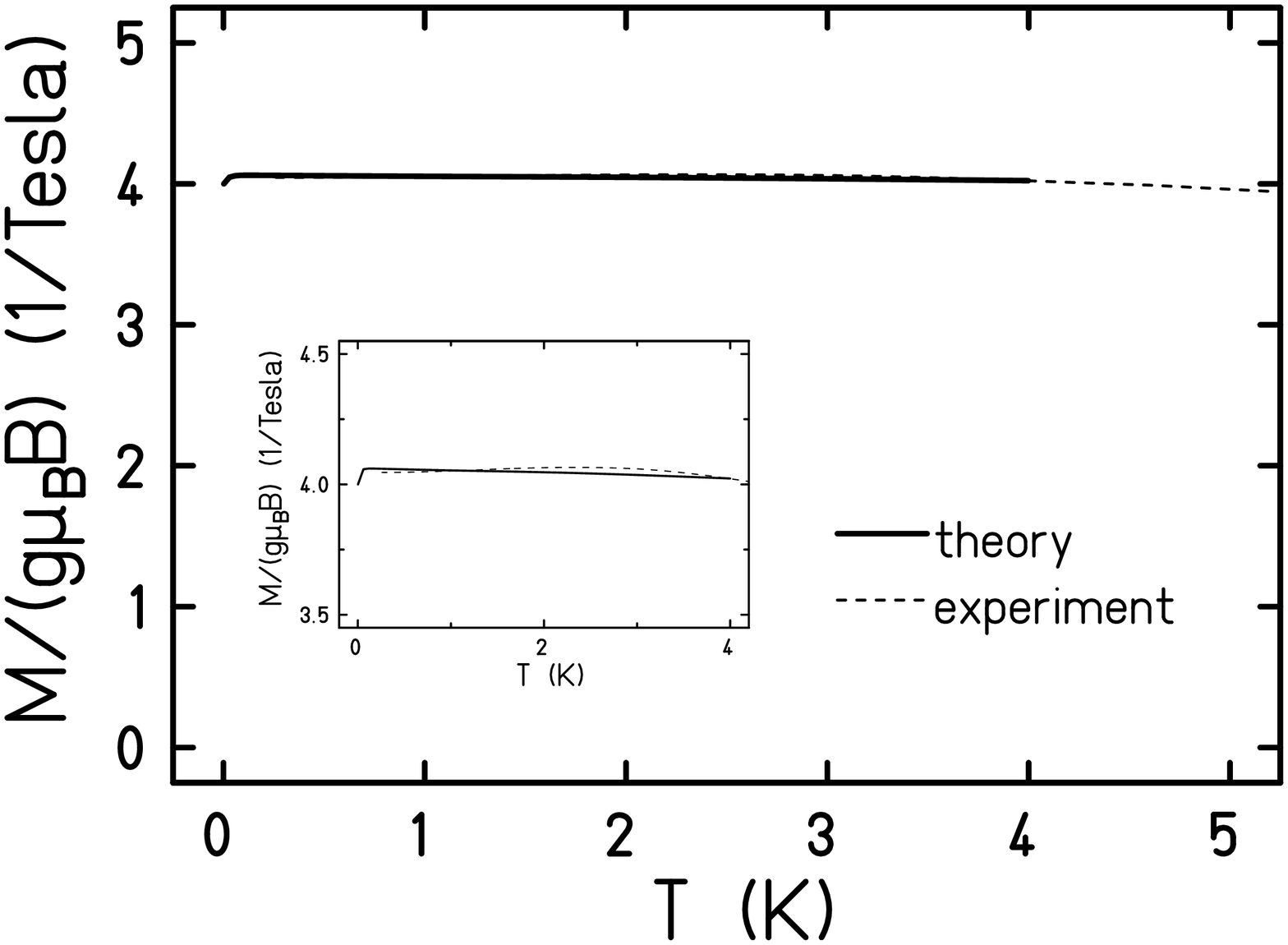,width=62mm}
$\quad$
\epsfig{file=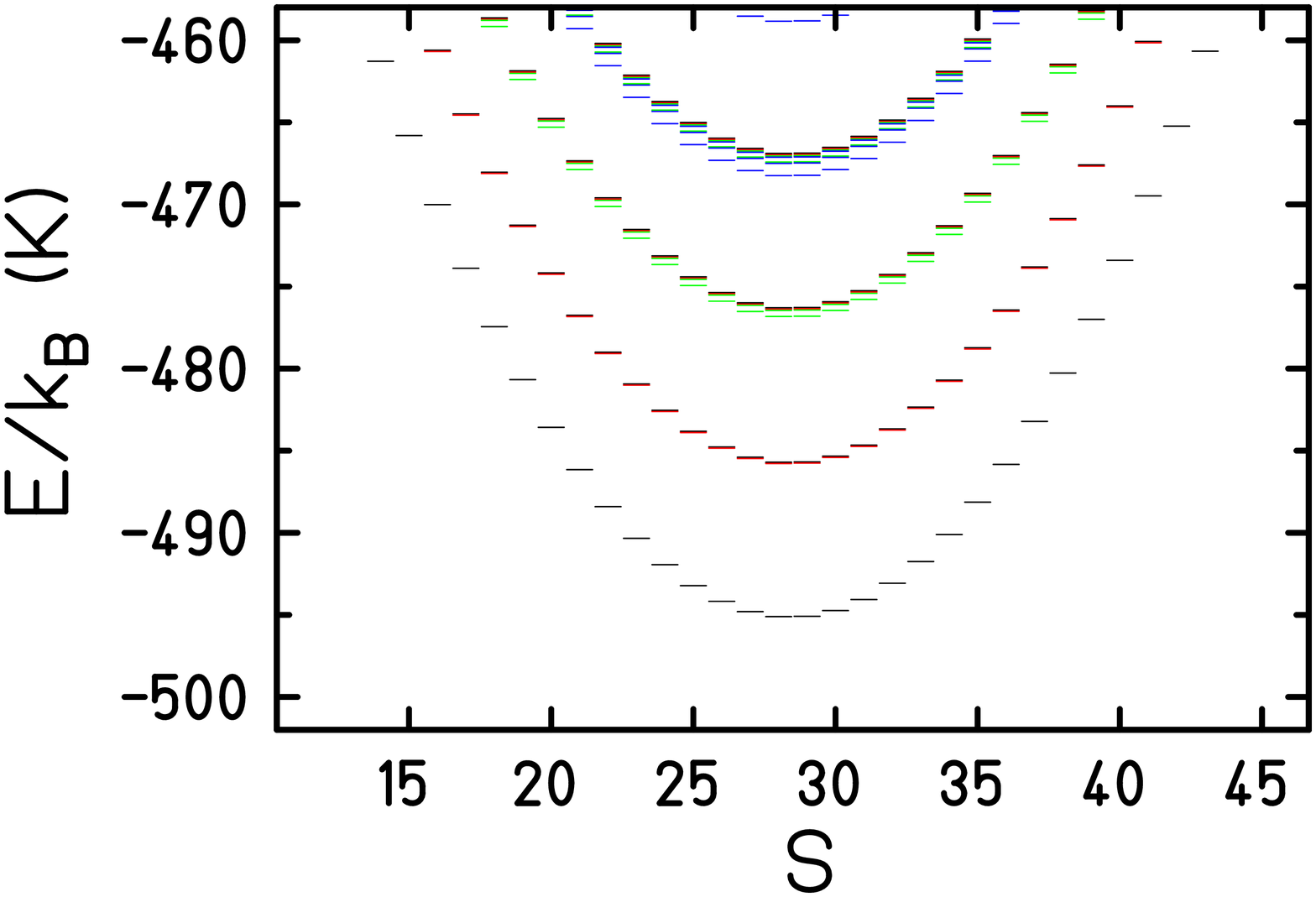,width=68mm}
\vspace*{1mm}
\caption[]{L.h.s.: Magnetisation vs. temperature using the four lowest
bands of the improved approximate Hamiltonian \fmref{E-3} with
$B=7.0$~T (solid line); experimental data (dashed line). 
The near constancy of ${\mathcal M}$ with
$T$ provides strong evidence that the lowest
rotational bands are indeed parabolic.
R.h.s.:
Low-lying energy eigenvalues in the range shown using
$\op{H}_2^{\text{eff}}$ with $B=7$~T.
}
\label{F-3}
\end{center} 
\end{figure} 
We now compare our theoretical results using \fmref{E-3} and a
Zeeman term to our magnetisation measurements at fixed magnetic
fields.  For both measured field strengths $B=5.5$~T (not shown)
and $B=7.0$~T (\figref{F-3}, l.h.s.) we find good agreement to
the experimental data.  The seemingly unexciting result, that
both the experimental and theoretical magnetisation M are
virtually constant over a wide temperature range, actually
provides important, indirect confirmation of the existence of a
sequence of rotational bands. This can well be understood with
the help of the spectrum of $\op{H}_2^{\text{eff}}$ shown in
\figref{F-3}~(r.h.s.) for a finite magnetic field. The levels of
the lowest parabola in \figref{F-3}~(r.h.s.) originate from the
$M=-S$ levels of the lowest rotational band for $B=0$
(\figref{F-1}). Those of the second parabola in
\figref{F-3}~(r.h.s.) originate from the $M=-S+1$ levels of the
lowest rotational band as well as the $M=-S$ levels of the
second rotational band.  Note that the levels of the parabolas
are distributed in a very symmetrical manner about their common
minimizing value of $S$ for the given $B$. Therefore these
levels are populated symmetrically and this leads to a
magnetisation which is nearly temperature independent. A slight
temperature dependence for $T\geap 2$~K (see inset of
\figref{F-3}, l.h.s.) stems from small differences in the
degeneracies of the individual levels as well as from the fact
that the approximate spectrum deviates from the exact one for
higher excitation energies. Nevertheless, the near constancy of
${\mathcal M}$ with $T$ is therefore also strong evidence that
not only the lowest rotational band \fmref{E-5}, but also the
second is parabolic, for otherwise the symmetry would be broken
and ${\mathcal M}$ would vary significantly with $T$.

\begin{figure}[ht!]
\begin{center}
\unitlength1mm
\begin{picture}(130,60)
\put(0, 0){\epsfig{file=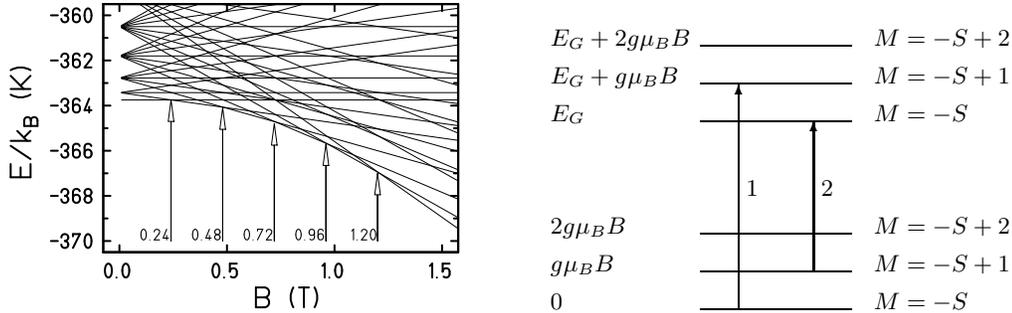,width=60mm}}
\put( 97, 0){\vector(0,1){30}}
\put( 98,15){$1$}
\put(107, 5){\vector(0,1){20}}
\put(108,15){$2$}
\put( 72, 0){$0$}
\put( 92, 0){\line(1,0){20}}
\put(115, 0){$M=-S$}
\put( 72, 5){$g\mu_B B$}
\put( 92, 5){\line(1,0){20}}
\put(115, 5){$M=-S+1$}
\put( 72,10){$2g\mu_B B$}
\put( 92,10){\line(1,0){20}}
\put(115,10){$M=-S+2$}
\put( 72,25){$E_G$}
\put( 92,25){\line(1,0){20}}
\put(115,25){$M=-S$}
\put( 72,30){$E_G+g\mu_B B$}
\put( 92,30){\line(1,0){20}}
\put(115,30){$M=-S+1$}
\put( 72,35){$E_G+2g\mu_B B$}
\put( 92,35){\line(1,0){20}}
\put(115,35){$M=-S+2$}
\end{picture}
\vspace*{1mm}
\caption[]{L.h.s.:Energy levels of the lowest rotational band as
functions of magnetic field.  R.h.s. Low-temperature
transitions, labelled 1 and 2, which will give rise to peaks in
an EPR spectrum involving the lowest Zeeman levels of the two
lowest rotational bands, separated by an energy gap
$E_G$. The gap size $E_G$ is independent of $S$, because the
rotational bands are parallel.}
\label{F-5}
\end{center} 
\end{figure} 
The existence of rotational bands implies also the occurrence of
resonances of the spin-lattice relaxation rate at low
temperatures \cite{JJL:PRL99}.  Proton NMR probes whether two
levels are separated by $\hbar\omega_L^p$ where $\omega_L^p$ is
the proton Larmor frequency. Since this energy is small compared
to the spacings between successive levels of the rotational band
in the absence of a magnetic field, resonances of the relaxation
rate could be seen close to level crossings induced by the
static magnetic field \cite{JJL:PRL99}.  In the absence of a
crystal field the level crossings will be observable at the
magnetic fields highlighted in the left panel \figref{F-5}. The
zero-field level spacings of the lowest rotational band are
$\Delta_S=2(S+1)\cdot 0.16$~K, and thus it will be necessary to
go to very low temperatures in order to observe distinct
resonance peaks.

We now discuss some of the other consequences of the existence
of a sequence of rotational bands. Using Eqs.~\fmref{E-2} or
\fmref{E-3} we estimate that the gap between the lowest two
rotational bands is $\Delta\approx 8$~K.  Inelastic neutron
scattering and EPR techniques should provide useful tests of
this prediction.  In the standard EPR setup, working with a
fixed resonance frequency, $\nu_0$, and an applied field of
variable strength $B$, because of the selection rule $\Delta
S=0$ one can focus on the Zeeman levels of the lowest two
rotational bands for a fixed value of $S$.  In the right panel
of \figref{F-5} we illustrate two distinct cases where the
resonance condition is met for a given value of $B$. The arrow
marked $1$ applies for a case where the inequality $\nu_G <
\nu_0 < 2 \nu_G$ is satisfied, where $\nu_G=E_G/h$ is the gap
frequency associated with successive rotational bands.  When the
resonance condition $h \nu_0 = E_G+g\mu_B B$ is met (independent
of $S$) the additional selection rule $\Delta M= \pm 1$ allows
for a transition from the $M=-S$ level of the lowest rotational
band to the $M=-S+1$ level of the first excited rotational band.
For the arrow marked $2$, which illustrates the case $\nu_0 <
\nu_G$, the resonance condition reads $h \nu_0 = E_G-g\mu_B B$,
and it is associated with a transition from the $M=-S+1$ level
of the lowest rotational band to the $M=-S$ level of the first
excited rotational band. The intensity of the peak will be
temperature dependent, proportional to the Boltzmann factor
$\exp(-\beta g \mu_B B)$.

Anticipating that the Debye temperature of
\{Mo$_{72}$Fe$_{30}$\} is of order 200--300~K, we believe that
specific heat measurements in the temperature range below 0.75~K
can serve as a useful probe of the levels of the lowest
rotational band. A comparison between our theory predictions and
specific heat data will be given elsewhere \cite{BCL}.

\section{Summary}

In summary, in this Letter we have presented an approximate
quantum model enabling us to determine the low-temperature
properties of the molecular magnet \{Mo$_{72}$Fe$_{30}$\}.
Several of our results have been confirmed by our measurements
and others remain to be tested.  The form of the effective
Hamiltonian, Eqs. \fmref{E-2} or \fmref{E-3}, is set by
incorporating the symmetry properties of the spin array which
leads to an identification of interacting sublattice spin
vectors. Some features of the model (e.g., parallel rotational
bands also for high-lying states) are oversimplified, however,
these do not play any role on results shown here. The present
method, which is based on the notion of parallel rotational
excitation bands, offers an insightful and quantitatively useful
platform as an alternative to the insurmountable difficulties in
treating the exact quantum Heisenberg model. Adaptation of this
methodology to still larger, high-symmetry magnetic molecules
that will be synthesized in the future can be expected to
provide similar benefits.

\acknowledgments We thank A. M\"uller (Bielefeld) and members of
his group, especially P.~K\"ogerler, both for providing us with
samples of \{Mo$_{72}$Fe$_{30}$\} as well as helpful
discussions. We also thank F.~Borsa, S.~Bud'ko, P.~Canfield,
M.~Exler, H.-J.~Schmidt, C.~Schr\"oder, and J.~Shinar for many
helpful discussions.  Finally we thank the National Science
Foundation and the Deutscher Akademischer Austauschdienst for
supporting a mutual exchange program.  The Ames Laboratory is
operated for the United States Department of Energy by Iowa
State University under Contract No. W-7405-Eng-82.

\end{document}